\documentclass{pnastwo}

\usepackage[dvips]{graphicx}
\usepackage{PNAStwoF}
\usepackage{amssymb,amsfonts,amsmath}
\usepackage{epstopdf}

\def\E{{\mathbb E}}

\def\ux{\underline{x}}

\def\cX{{\cal X}}
\def\cS{{\cal S}}
\def\cN{{\cal N}}
\def\elb{\bar{\ell}}
\def\oeta{\overline{\eta}}
\def\onu{\overline{\nu}}
\def\E{\mathbb{E}}
\def\ux{\underline{x}}

\def\Z{{\cal Z}}
\def\uxb{\underline{x}_{\mbox{\tiny B}}}
\def\bo{{\rm B}}
\def\de{{\rm d}}

\def\da{\partial a}
\def\di{\partial i}

\begin{document}

\conflictofinterest{Conflict of interest footnote placeholder}


\footcomment{Abbreviations: rCSP, random constraint satisfaction
problem; 1RSB, one-step replica symmetry breaking}

\title{Gibbs states and the set of solutions of random constraint
satisfaction problems}

\author{Florent Krzakala\affil{1}{Lab.~PCT, ESPCI, Paris, France}, 
Andrea Montanari\affil{2}{Depts of Electrical Engineering and
  Statistics, Stanford University, USA}\affil{3}{LPTENS, Ecole Normale
  Sup\'{e}rieure and UPMC, Paris, France}\thanks{To whom
  correspondence should be addressed. E-mail: montanari@stanford.edu},
Federico Ricci-Tersenghi\affil{4}{Dipartimento di Fisica and CNR-INFM,
 Universit\`{a} ``La Sapienza'', Roma, Italy},
Guilhem Semerjian\affil{3}{},\and
Lenka Zdeborov\'a\affil{5}{LPTMS, Universit\'e de Paris-Sud, Orsay,
  France}}

\contributor{Submitted to Proceedings of the National Academy of Sciences 
of the United States of America}

\maketitle

\begin{article}

\begin{abstract}  
An instance of a random constraint satisfaction problem defines a
random subset $\cS$ (the set of solutions) of a large product space
$\cX^N$ (the set of assignments).  We consider two prototypical
problem ensembles (random $k$-satisfiability and $q$-coloring of
random regular graphs), and study the uniform measure with support on
$\cS$.  As the number of constraints per variable increases, this
measure first decomposes into an exponential number of pure states
(`clusters'), and subsequently condensates over the largest such
states. Above the condensation point, the mass carried by the $n$
largest states follows a Poisson-Dirichlet process.

For typical large instances, the two transitions are sharp.  We
determine for the first time their precise location.  Further, we
provide a formal definition of each phase transition in terms of
different notions of correlation between distinct variables in the
problem.

The degree of correlation naturally affects the performances of many
search/sampling algorithms. Empirical evidence suggests that local
Monte Carlo Markov Chain strategies are effective up to the clustering
phase transition, and belief propagation up to the condensation point.
Finally, refined message passing techniques (such as survey
propagation) may beat also this threshold.
\end{abstract}

%
%


\dropcap{C}onstraint satisfaction problems (CSPs) arise in a large
spectrum of scientific disciplines.  An instance of a CSP is said
to be satisfiable if there exists an assignment
of $N$ variables $(x_1,x_2,\dots,x_N)\equiv \ux$, $x_i\in\cX$ ($\cX$
being a finite alphabet) which satisfies all the constraints within
a given collection. The problem consists in finding such an assignment
or show that the constraints are unsatisfiable.
More precisely, one is given a set of functions
$\psi_a:\cX^k\to \{0,1\}$, with $a\in \{1,\dots,M\}\equiv [M]$ and of
$k$-tuples of indices $\{i_a(1),\dots, i_a(k)\}\subseteq [N]$, and has
to establish whether there exists $\ux\in\cX^N$ such that
$\psi_a(x_{i_a(1)},\dots,x_{i_a(k)})=1$ for all $a$'s. In this article
we shall consider two well known families of CSP's (both known to be
NP-complete \cite{GaJo}):
\begin{enumerate} 
\item[$(i)$] $k$-satisfiability ($k$-SAT) with $k\ge 3$. In this case
$\cX=\{0,1\}$.  The constraints are defined by fixing a $k$-tuple
$(z_a(1),\dots,z_a(k))$ for each $a$, and setting
$\psi_a(x_{i_a(1)},\dots,x_{i_a(k)})=0$ if
$(x_{i_a(1)},\dots,x_{i_a(k)})=(z_a(1),\dots,z_a(k))$ and $=1$
otherwise.
\item[$(ii)$] $q$-coloring ($q$-COL) with $q\ge 3$. Given a graph $G$
with $N$ vertices and $M$ edges, one is asked to assign colors
$x_i\in\cX\equiv \{1,\dots,q\}$ to the vertices in such a way that no
edge has the same color at both ends.
\end{enumerate}

\begin{figure}[t]
\centerline{\includegraphics[width=0.35\linewidth]{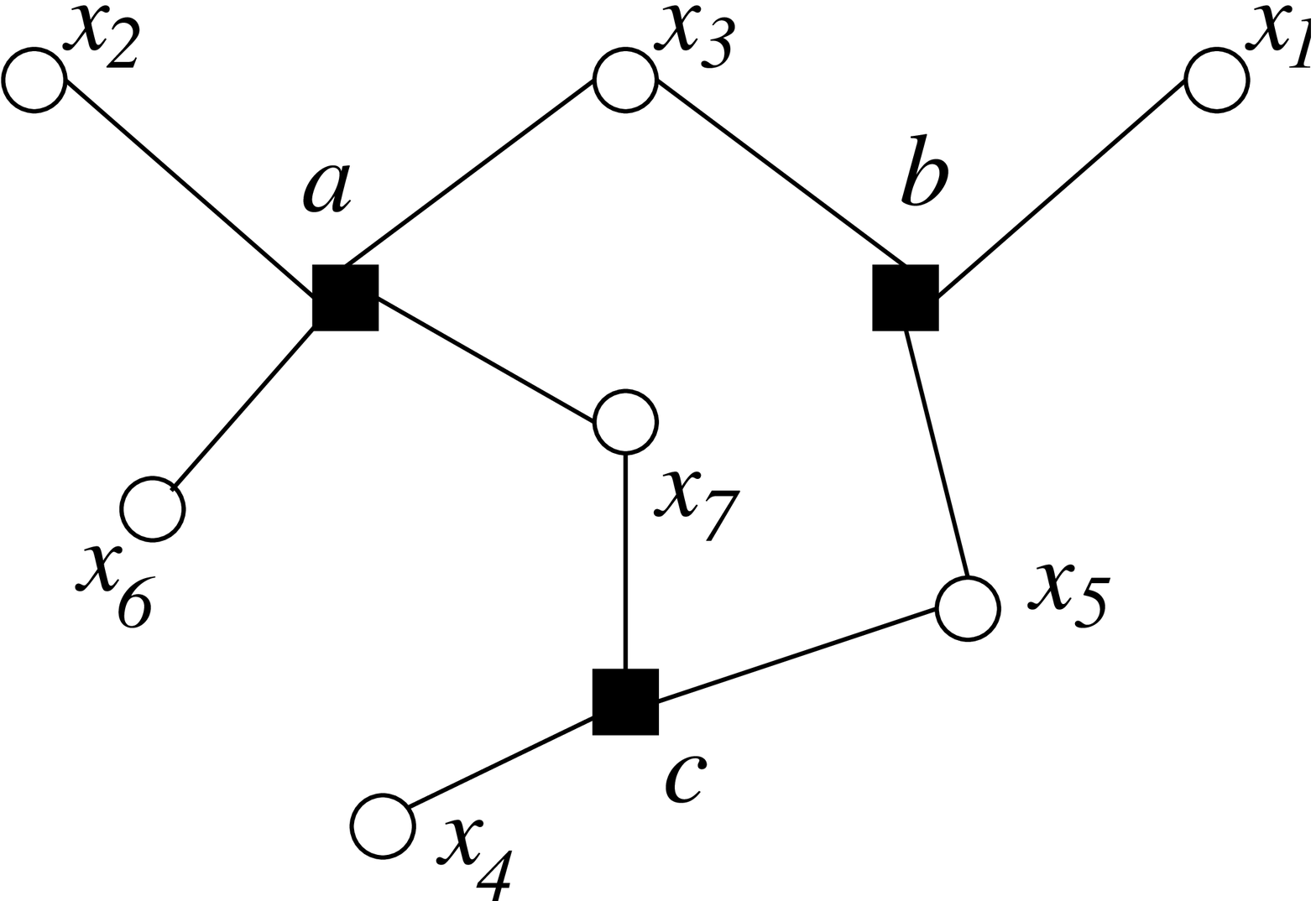}}
\caption{The factor graph of a small CSP allows to define the distance
  $d(i,j)$ between variables $x_i$ and $x_j$ (filled squares are
  constraints and empty circles variables). Here, for instance,
  $d(6,1) = 2$ and $d(3,5)=1$.\label{fig:Factor}}
\end{figure}

The optimization (maximize the number of satisfied constraints) and
counting (count the number of satisfying assignments) versions of this
problems are defined straightforwardly.  It is also convenient to
represent CSP instances as factor graphs \cite{Factor}, i.e.\
bipartite graphs with vertex sets $[N]$, $[M]$ including an edge
between node $i\in[N]$ and $a\in[M]$ if and only if the $i$-th
variable is involved in the $a$-th constraint,
cf.~Fig.~\ref{fig:Factor}.  This representation allows to define
naturally a distance $d(i,j)$ between variable nodes.

Ensembles of random CSP's (rCSP) were introduced (see
e.g.~\cite{FrancoPaull}) with the hope of discovering generic
mathematical phenomena that could be exploited in the design of
efficient algorithms.  Indeed several search heuristics, such as
Walk-SAT \cite{WalkSAT} and `myopic' algorithms \cite{Myopic} have
been successfully analyzed and optimized over rCSP ensembles.  The
most spectacular advance in this direction has probably been the
introduction of a new and powerful message passing algorithm (`survey
propagation', SP) \cite{MarcGiorgioRiccardo}.  The original
justification for SP was based on the (non-rigorous) cavity method
from spin glass theory.  Subsequent work proved that standard message
passing algorithms (such as belief propagation, BP) can indeed be
useful for some CSP's
\cite{BayatiEtAl,GamarnikColoring,MontanariShah}.  Nevertheless, the
fundamental reason for the (empirical) superiority of SP in this
context remains to be understood and a major open problem in the
field. Building on a refined picture of the solution set of rCSP, this
paper provides a possible (and testable) explanation.
We consider two ensembles that have attracted the majority of work in
the field: $(i)$ random $k$-SAT: each $k$-SAT instance with $N$
variables and $M=N\alpha$ clauses is considered with the same
probability; $(ii)$ $q$-COL on random graphs: the graph $G$ is
uniformly random among the ones over $N$ vertices, with uniform degree
$l$ (the number of constraints is therefore $M=Nl/2$).

%
%
\subsection{Phase transitions in random CSP}

It is well known that rCSP's may undergo phase transitions as the
number of constraints per variable $\alpha$ is varied\footnote{For
coloring $l$-regular graphs, we can use $l = 2\alpha$ as a
parameter. When considering a phase transition defined through some
property ${\cal P}$ increasing in $l$, we adopt the convention of
denoting its location through the smallest integer such that ${\cal
P}$ holds.}.  The best known of such phase transitions is the
SAT-UNSAT one: as $\alpha$ crosses a critical value $\alpha_{\rm
s}(k)$ (that can, in principle, depend on $N$), the instances pass
from being satisfiable to unsatisfiable with high
probability\footnote{The term `with high probability' (whp) means with
probability approaching one as $N\to\infty$.}  \cite{Friedgut}.  For
$k$-SAT, it is known that $\alpha_{\rm s}(2)=1$. A conjecture based on
the cavity method was put forward in \cite{MarcGiorgioRiccardo} for
all $k\ge 3$ that implied in particular the values presented in
Table~\ref{Table:crit} and $\alpha_{\rm s}(k) = 2^k\log 2
-\frac{1}{2}(1+\log 2)+O(2^{-k})$ for large $k$ \cite{Mertens}.
Subsequently it was proved that $\alpha_{\rm s}(k) \ge 2^k\log 2-O(k)$
confirming this asymptotic behavior \cite{AchlioptasNaorPeres}. An
analogous conjecture for $q$-coloring was proposed in
\cite{ColTrieste} yielding, for regular random graphs
\cite{KrzakalaPagnaniWeigt}, the values reported in
Table~\ref{Table:crit} and $l_{\rm s}(q) = 2q\log{q} -\log{q}-1 +
o(1)$ for large $q$ (according to our convention, random graphs are
whp uncolorable if $l\ge l_{\rm s}(q)$). It was proved in
\cite{Luczak,AchlioptasNaorPeres} that $l_{\rm s}(q) = 2q\log q -
O(\log q)$.

Even more interesting and challenging are phase transitions in the
structure of the set $\cS\subseteq\cX^N$ of solutions of rCSP's
(`structural' phase transitions). Assuming the existence of solutions,
a convenient way of describing $\cS$ is to introduce the uniform
measure over solutions $\mu(\ux)$:
\begin{eqnarray}
\mu(\ux) = \frac{1}{Z}\, \prod_{a=1}^M\psi_a(x_{i_a(1)},\dots,x_{i_a(k)})
\, ,\label{eq:Measure}
\end{eqnarray}
where $Z\ge 1$ is the number of solutions. Let us stress that, since
$\cS$ depends on the rCSP instance, $\mu(\,\cdot\,)$ is itself random.

\begin{table}[t]
\caption{Critical connectivities for the dynamical, condensation and
  satisfiability transitions in k-SAT and q-COL\label{Table:crit}}
\begin{tabular}{lrrrlrrr}
\hline
SAT & $\alpha_{\rm d}$ & $\alpha_{\rm c}$ &
$\alpha_{\rm s}$\cite{Mertens} & COL &
$l_{\rm d}$\cite{MontanariMezard} & $l_{\rm c}$ & 
$l_{\rm s}$\cite{KrzakalaPagnaniWeigt}\\
\hline
$k=4$ & $9.38 $ & $9.547$ & $9.93$ & $q=4$ & $9$ & $10$ & $10$ \\
$k=5$ & $19.16$ & $20.80$ & $21.12$ & $q=5$ & $14$ & $14$ & $15$ \\
$k=6$ & $36.53$ & $43.08$ & $43.4$ & $q=6$ & $18$ & $19$ & $20$ \\
\hline
\end{tabular}
\end{table}

We shall now introduce a few possible `global' characterizations of
the measure $\mu(\,\cdot\,)$. Each one of these properties has its
counterpart in the theory of Gibbs measures and we shall partially
adopt that terminology here \cite{Georgii}.

In order to define the first of such characterizations, we let $i\in
[N]$ be a uniformly random variable index, denote as $\ux_{\ell}$ the
vector of variables whose distance from $i$ is at least $\ell$, and by
$\mu(x_i|\ux_{\ell})$ the marginal distribution of $x_i$ given
$\ux_{\ell}$.  Then we say that the measure \eqref{eq:Measure}
satisfies the uniqueness condition if, for any given $i\in[N]$,
\begin{eqnarray}
\E\sup_{\ux_\ell,\ux'_{\ell}}
\sum_{x_i\in\cX}\left|\mu(x_i|\ux_{\ell}) - \mu(x_i|\ux'_{\ell})\right|
\to 0\, .\label{eq:Uniqueness}
\end{eqnarray}
as $\ell\to \infty$ (here and below the limit $N\to\infty$ is
understood to be taken before $\ell\to\infty$).  This expresses a
`worst case' correlation decay condition. Roughly speaking: the
variable $x_i$ is (almost) independent of the far apart variables $\ux_{\ell}$
{\em irrespective} is the instance realization and the variables
distribution outside the horizon of radius $\ell$. The threshold for
uniqueness (above which uniqueness ceases to hold) was estimated in
\cite{MontanariShah} for random $k$-SAT, yielding $\alpha_{\rm u}(k) =
(2\log k)/k[1+o(1)]$ (which is asymptotically close to the threshold
for the pure literal heuristics) and in \cite{Jonasson} for coloring
implying $l_{\rm u}(q) = q$ for $q$ large enough
(a `numerical' proof of the same statement exists for small $q$).  Below such
thresholds BP can be proved to return good estimates of the local
marginals of the distribution \eqref{eq:Measure}.

Notice that the uniqueness threshold is far below the SAT-UNSAT
threshold. Furthermore, several empirical studies
\cite{MPEmpirical,Maneva} pointed out that BP (as well as many other
heuristics \cite{WalkSAT,Myopic}) is effective up to much larger
values of the clause density. In a remarkable series of papers
\cite{BiroliMonassonWeigt,MarcGiorgioRiccardo}, statistical physicists
argued that a second structural phase transition is more relevant than
the uniqueness one. Following this literature, we shall refer to this
as the `dynamic phase transition' (DPT) and denote the corresponding
threshold as $\alpha_{\rm d}(k)$ (or $l_{\rm d}(q)$).  In order to
precise this notion, we provide here two alternative formulations
corresponding to two distinct intuitions.  According to the first one,
above $\alpha_{\rm d}(k)$ the variables $(x_1,\dots,x_N)$ become
globally correlated under $\mu(\,\cdot\,)$. The criterion in
\eqref{eq:Uniqueness} is replaced by one in which far apart variables
$\ux_{\ell}$ are themselves sampled from $\mu$ (`extremality'
condition):
\begin{eqnarray}
\E\sum_{\ux_{\ell}} \mu(\ux_{\ell})
\sum_{x_i}\left|\mu(x_i|\ux_{\ell}) - \mu(x_i)\right|
\to 0\, .\label{eq:Correlation}
\end{eqnarray}
as $\ell\to\infty$.  The infimum value of $\alpha$ (respectively $l$)
such that this condition is no longer fulfilled is the threshold
$\alpha_{\rm d}(k)$ ($l_{\rm d}(k)$). Of course this criterion is
weaker than the uniqueness one (hence $\alpha_{\rm d}(k)\ge\alpha_{\rm
u}(k)$).

According to the second intuition, above $\alpha_{\rm d}(k)$, the
measure \eqref{eq:Measure} decomposes into a large number of
disconnected `clusters'. This means that there exists a partition
$\{A_{n}\}_{n=1\dots \cN}$ of $\cX^N$ (depending on the instance) such
that: $(i)$ One cannot find $n$ such that $\mu(A_{n})\to 1$; $(ii)$
Denoting by $\partial_\epsilon A$ the set of configurations $\ux\in
\cX^N\backslash A$ whose Hamming distance from $A$ is at most
$N\epsilon$, we have $\mu(\partial_\epsilon
A_{n})/\mu(A_n)(1-\mu(A_n))\to 0$ exponentially fast in $N$ for all
$n$ and $\epsilon$ small enough. Notice that the measure $\mu$ can be
decomposed as
\begin{eqnarray}
\mu(\,\cdot\,) = \sum_{n=1}^{\cN} w_n\, \mu_n(\,\cdot\, )\, ,
\label{eq:Decomposition}
\end{eqnarray}
where $w_n \equiv \mu(A_n)$ and
$\mu_n(\,\cdot\,)\equiv\mu(\,\cdot\,|A_n)$.  We shall always refer to
$\{ A_n\}$ as the `finer' partition with these properties.

The above ideas are obviously related to the performance of
algorithms.  For instance, the correlation decay condition in
\eqref{eq:Correlation} is likely to be sufficient for approximate
correctness of BP on random formulae.  Also, the existence of
partitions as above implies exponential slowing down in a large class
of MCMC sampling algorithms\footnote{One possible approach to the
definition of a MCMC algorithm is to relax the constraints by setting
$\psi_a(\cdots)=\epsilon$ instead of $0$ whenever the $a$-th
constraint is violated. Glauber dynamics can then be used to sample
from the relaxed measure $\mu_{\epsilon}(\,\cdot\,)$.}.

Recently, some important rigorous results were obtained supporting
this picture~\cite{MezardMoraZecchina,AchlioptasRicci}.  However, even
at the heuristic level, several crucial questions remain open. The
most important concern the distribution of the weights $\{w_n\}$: are
they tightly concentrated (on an appropriate scale) or not? A
(somewhat surprisingly) related question is: can the absence of
decorrelation above $\alpha_{\rm d}(k)$ be detected by probing a
subset of variables bounded in $N$?

SP \cite{MarcGiorgioRiccardo} can be thought as an inference algorithm
for a modified graphical model that gives unit weight to each cluster
\cite{AlfredoRiccardo,Maneva}, thus tilting the original measure
towards small clusters. The resulting performances will strongly
depend on the distribution of the cluster sizes $w_n$.  Further, under
the tilted measure, $\alpha_{\rm d}(k)$ is underestimated because
small clusters have a larger impact.  The correct value was never
determined (but see \cite{MontanariMezard} for coloring).  The authors
of \cite{PalassiniMezardRivoire} undertook the technically challenging
task of determining the cluster size distribution, without however
clarifying several of its properties.

In this paper we address these issues, and unveil at least two
unexpected phenomena. Our results are described in the next Section
with a summary just below.  Finally we will discuss the connection
with the performances of SP.  Some technical details of the
calculation are collected in the last Section.
%
%
\section{Results and discussion}

The formulation in terms of extremality condition,
cf.~Eq.~\eqref{eq:Correlation}, allows for an heuristic calculation of
the dynamic threshold $\alpha_{\rm d}(k)$. Previous attempts were
based instead on the cavity method, that is an heuristic
implementation of the definition in terms of pure state decomposition,
cf.~Eq.~\eqref{eq:Decomposition}. Generalizing the results of
\cite{MontanariMezard}, it is possible to show that the two
calculations provide identical results. However, the first one is
technically simpler and under much better control. As mentioned above
we obtain, for all $k\ge 4$ a value of $\alpha_{\rm d}(k)$ larger than
the one quoted in \cite{MarcGiorgioRiccardo,Mertens}.

Further we determined the distribution of cluster sizes $w_n$, thus
unveiling a third `condensation' phase transition at $\alpha_{\rm
c}(k)\ge \alpha_{\rm d}(k)$ (strict inequality holds for $k\ge 4$ in
SAT and $q\ge 4$ in coloring, see below).  For $\alpha<\alpha_{\rm
c}(k)$ the weights $w_n$ concentrate on a logarithmic scale (namely
$-\log w_n$ is $\Theta(N)$ with $\Theta(N^{1/2})$
fluctuations). Roughly speaking the measure is evenly split among an
exponential number of clusters.

\begin{figure}[t]
\center{\includegraphics[width=0.95\linewidth]{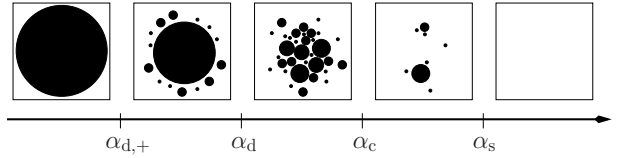}}
\caption{Pictorial representation of the different phase transitions
in the set of solutions of a rCSP. At $\alpha_{{\rm d},+}$ some
clusters appear, but for $\alpha_{{\rm d},+}<\alpha<\alpha_{\rm d}$
they comprise only an exponentially small fraction of solutions.  For
$\alpha_{\rm d}<\alpha<\alpha_{\rm c}$ the solutions are split among
about $e^{N\Sigma_*}$ clusters of size $e^{Ns_*}$. If $\alpha_{\rm
c}<\alpha<\alpha_{\rm s}$ the set of solutions is dominated by a few
large clusters (with strongly fluctuating weights), and above
$\alpha_{\rm s}$ the problem does not admit solutions any more.}
\end{figure}
\begin{figure}[t]
\center{\includegraphics[width=0.8\linewidth]{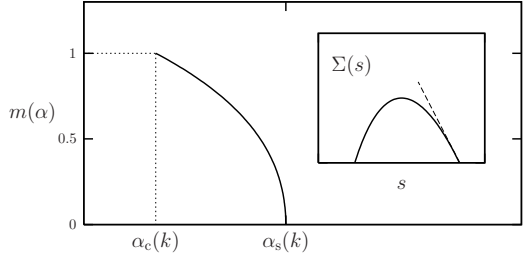}}
\caption{The Parisi 1RSB parameter $m(\alpha)$ as a function of the
constraint density $\alpha$.  In the inset, the complexity $\Sigma(s)$
as a function of the cluster entropy for $\alpha = \alpha_{\rm s}(k)-
0.1$ (the slope at $\Sigma(s) = 0$ is $-m(\alpha)$).  Both curves have
been computed from the large $k$ expansion.\label{fig:MofAlpha}}
\end{figure}

For $\alpha>\alpha_{\rm c}(k)$ (and $<\alpha_{\rm s}(k)$) the measure
is carried by a subexponential number of clusters.  More precisely,
the ordered sequence $\{w_n\}$ converges to a well known
Poisson-Dirichlet process $\{w^*_n\}$, first recognized in the spin
glass context by Ruelle \cite{Ruelle}.  This is defined by $w^*_n =
x_n/\sum x_n$, where $x_n>0$ are the points of a Poisson process with
rate $x^{-1-m(\alpha)}$ and $m(\alpha)\in (0,1)$.  This picture is
known in spin glass theory as `one step replica symmetry breaking'
(1RSB) and has been proven in Ref.~\cite{TalagrandPspin} for some
special models.  The `Parisi 1RSB parameter' $m(\alpha)$ is
monotonically decreasing from $1$ to $0$ when $\alpha$ increases from
$\alpha_{\rm c}(k)$, to $\alpha_{\rm s}(k)$,
cf.~Fig.~\ref{fig:MofAlpha}.

Remarkably the condensation phase transition is also linked to an
appropriate notion of correlation decay.  If $i(1),\dots,i(n)\in[N]$
are uniformly random variable indices, then, for $\alpha<\alpha_{\rm
c}(k)$ and any fixed $n$:
\begin{eqnarray}
\E\sum_{\{x_{i(\cdot)}\}} \left|\mu(x_{i(1)}\dots
x_{i(n)})-\mu(x_{i(1)})\cdots\mu(x_{i(n)})\right| \to 0
\label{eq:Correlation2}
\end{eqnarray}
as $N\to \infty$. Conversely, the quantity on the left hand side 
remains positive for
$\alpha>\alpha_{\rm c}(k)$.  It is easy to understand that this
condition is even weaker than the extremality one,
cf.~Eq.~\eqref{eq:Correlation}, in that we probe correlations of
finite subsets of the variables. In the next two Sections we discuss
the calculation of $\alpha_{\rm d}$ and $\alpha_{\rm c}$.
%
%
%
\subsection{Dynamic phase transition and Gibbs measure extremality}
\label{sec:Dynamical}

A rigorous calculation of $\alpha_{\rm d}(k)$ along any of the two
definitions provided above, cf.~Eqs.~\eqref{eq:Correlation} and
\eqref{eq:Decomposition} remains an open problem. Each of the two
approaches has however an heuristic implementation that we shall now
describe.  It can be proved that the two calculations yield equal
results as further discussed in the last Section of the paper.

The approach based on the extremality condition in
\eqref{eq:Correlation} relies on an easy-to-state assumption, and
typically provides a more precise estimate. We begin by observing
that, due to the Markov structure of $\mu(\,\cdot\, )$, it is
sufficient for Eq.~\eqref{eq:Correlation} to hold that the same
condition is verified by the correlation between $x_i$ and the set of
variables at distance exactly $\ell$ from $i$, that we shall keep
denoting as $\ux_{\ell}$.  The idea is then to consider a large yet
finite neighborhood of $i$.  Given $\elb\ge \ell$, the factor graph
neighborhood of radius $\elb$ around $i$ converges in distribution to
the radius--$\elb$ neighborhood of the root in a well defined random
tree factor graph $T$.

For coloring of random regular graphs, the correct limiting tree model
$T$ is coloring on the infinite $l$-regular tree.  For random $k$-SAT,
$T$ is defined by the following construction. Start from the root
variable node and connect it to $l$ new function nodes (clauses), $l$
being a Poisson random variable of mean $k\alpha$. Connect each of
these function nodes with $k-1$ new variables and repeat. The
resulting tree is infinite with non-vanishing probability if
$\alpha>1/k(k-1)$. Associate a formula to this graph in the usual way,
with each variable occurrence being negated independently with
probability $1/2$.

The basic assumption within the first approach is that the extremality
condition in \eqref{eq:Correlation} can be checked on the correlation
between the root and generation-$\ell$ variables in the tree model.
On the tree, $\mu(\,\cdot\,)$ is defined to be a translation invariant
Gibbs measure \cite{Georgii} associated to the infinite factor
graph\footnote{More precisely $\mu(\,\cdot\,)$ is obtained as a limit
of free boundary measures (further details in \cite{OurGibbs}).}  $T$
(which provides a specification).  The correlation between the root
and generation-$\ell$ variables can be computed through a recursive
procedure (defining a sequence of distributions $\overline{P}_{\ell}$,
see Eq.~\eqref{eq:M1Recursion} below). The recursion can be
efficiently implemented numerically yielding the values presented in
Table~\ref{Table:crit} for $k$ ({\it resp}. $q$)$=4,5,6$. For large
$k$ ({\it resp}. $q$) one can formally expand the equations on
$P_{\ell}$ and obtain
\begin{eqnarray}
\alpha_{\rm d}(k) &=& \frac{2^{k}}{k}\left[\log k+\log\log
k+\gamma_{\rm d}+ O\left(\frac{\log\log k}{\log k}\right)\right]\,
\label{eq:LargeKDyn} \\
l_{\rm d}(q) &=& q\ \left[ \log q + \log \log q +\gamma_{\rm d} +
o(1)\right] \label{eq:LargeKDynCol}
\end{eqnarray}
with $\gamma_{\rm d}=1$ (under a technical assumption on the
structure of $P_\ell$).

The second approach to the determination of $\alpha_{\rm d}(k)$ is
based on the `cavity method'
\cite{MarcGiorgioRiccardo,PalassiniMezardRivoire}. It begins by
assuming a decomposition in pure states of the form
\eqref{eq:Decomposition} with two crucial properties: $(i)$ If we
denote by $W_n$ the size of the $n$-th cluster (and hence
$w_n=W_n/\sum W_n$), then the number of clusters of size $W_n= e^{Ns}$
grows approximately as $e^{N\Sigma(s)}$; $(ii)$ For each
single-cluster measure $\mu_n(\,\cdot\,)$, a correlation decay
condition of the form \eqref{eq:Correlation} holds.

The approach aims at determining the rate function $\Sigma(s)$,
`complexity': the result is expressed in terms of the solution of a
distributional fixed point equation.  For the sake of simplicity we
describe here the simplest possible scenario\footnote{The precise
picture depends on the value of $k$ ({\it resp}. $q$) and can be
somewhat more complicated.} resulting from such a calculation,
cf.~Fig.~\ref{fig:Complexity}.  For $\alpha<\alpha_{{\rm
d},-\infty}(k)$ the cavity fixed point equation does not admit any
solution: no clusters are present.  At $\alpha_{{\rm d},-\infty}(k)$ a
solution appears, eventually yielding, for $\alpha>\alpha_{{\rm d},+}$
a non-negative complexity $\Sigma(s)$ for some values of $s\in{\mathbb
R}_+$.  The maximum and minimum such values will be denoted by $s_{\rm
max}$ and $s_{\rm min}$.  At a strictly larger value $\alpha_{{\rm
d},0}(k)$, $\Sigma(s)$ develops a stationary point (local maximum). It
turns out that $\alpha_{{\rm d},0}(k)$ coincides with the threshold
computed in \cite{MarcGiorgioRiccardo,Mertens,KrzakalaPagnaniWeigt}.
In particular $\alpha_{{\rm d},0}(4)\approx 8.297$, $\alpha_{{\rm
d},0}(5)\approx 16.12$, $\alpha_{{\rm d},0}(6)\approx 30.50$ and
$l_{{\rm d},0}(4)=9$, $l_{{\rm d},0}(5)=13$, $l_{{\rm d},0}(6)=17$.
For large $k$ ({\it resp}. $q$), $\alpha_{{\rm d},0}(k)$ admits the
same expansion as in Eqs.~\eqref{eq:LargeKDyn},
\eqref{eq:LargeKDynCol} with $\gamma_{{\rm d},0} = 1-\log 2$.
However, up to the larger value $\alpha_{\rm d}(k)$, the appearance of
clusters is irrelevant from the point of view of $\mu(\,\cdot\,)$. In
fact, within the cavity method it can be shown that
$e^{N[s+\Sigma(s)]}$ remains exponentially smaller than the total
number of solutions $Z$: most of the solutions are in a single
``cluster''.  The value $\alpha_{\rm d}(k)$ is determined by the
appearance of a point $s_*$ with $\Sigma'(s_*) = -1$ on the complexity
curve.  Correspondingly, one has $Z\approx e^{N[\Sigma(s_*)+s_*]}$:
most of the solutions are comprised in clusters of size about
$e^{Ns_*}$.  The entropy per variable $\phi = \lim_{\substack N
\rightarrow \infty} N^{-1}\log Z$ remains analytic at $\alpha_{\rm
d}(k)$.
%
%
\subsection{Condensation phase transition}
\label{sec:Condensation}

\begin{figure}[t]
\center{\includegraphics[width=0.85\linewidth]{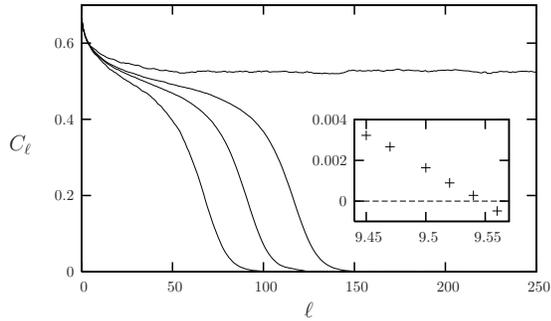}}
\caption{Correlation function \eqref{eq:Correlation} between the root
and generation $\ell$ variables in a random $k$-SAT tree formula. Here
$k=4$ and (from bottom to top) $\alpha = 9.30$, $9.33$, $9.35$, $9.40$
(recall that $\alpha_{\rm d}(4)\approx 9.38$). In the inset, the
complexity $\Sigma(s_*)$ of dominant clusters as a function of
$\alpha$ for $4$-SAT.\label{fig:Correlation}}
\end{figure}

As $\alpha$ increases above $\alpha_{\rm d}$, $\Sigma(s_*)$ decreases:
clusters of highly correlated solutions may no longer satisfy the
newly added constraints. In the inset of Fig.~\ref{fig:Correlation} we
show the $\alpha$ dependency of $\Sigma(s_*)$ for $4$-SAT. In the
large $k$ limit, with $\alpha=\rho\, 2^{k}$ we get $\Sigma(s^*) = \log
2-\rho-\log 2\, e^{-k\rho}+O(2^{-k})$, and $s_* = \log 2 \,
e^{-k\rho}+O(2^{-k})$.

\begin{figure}[t]
\center{\includegraphics[width=0.87\linewidth]{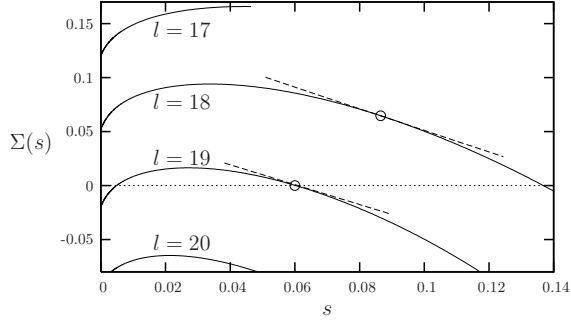}}
\caption{The complexity function (the number of clusters with entropy
density $s$ is $e^{N\Sigma(s)}$) for the $6$-colorings of $l$-regular
graphs with $l \in \{17,18,19,20\}$.  Circles indicate the dominating
states with entropy $s_*$; the dashed lines have slopes
$\Sigma'(s_*)=-1$ for $l=18$ and $\Sigma'(s_*)=-0.92$ for $l=19$.  The
dynamic phase transition is $l_{\rm d}(6)=18$, the condensation one
$l_{\rm d}(6)=19$, and the SAT-UNSAT one $l_{\rm s}(6)=20$.
\label{fig:Complexity}}
\end{figure}

The condensation point $\alpha_{\rm c}(k)$ is the value of $\alpha$
such that $\Sigma(s_*)$ vanishes: above $\alpha_{\rm c}(k)$, most of
the measure is contained in a sub-exponential number of large
clusters\footnote{Notice that for $q$-coloring, since $l$ is an
integer, the `condensated' regime $[l_{\rm c}(q),l_{\rm s}(q)]$ may be
empty: This is the case for $q$$=$$4$.  On the contrary, $q$$=$$5$ is
always condensated for $l_{\rm d}<l<l_{\rm s}$.}.  Our estimates for
$\alpha_{\rm c}(k)$ are presented in Table~\ref{Table:crit} (see also
Fig.~\ref{fig:Complexity} for $\Sigma(s)$ in the $6$-coloring) while
in the large-$k$ limit we obtain $\alpha_{\rm c}(k) =2^{k}\log 2 -
\frac{3}{2}\log 2+O(2^{-k})$ [recall that the SAT-UNSAT transition is
at $\alpha_{\rm s}(k) = 2^{k}\log 2-\frac{1+\log 2}{2}+O(2^{-k})$] and
$l_{\rm c}(q) =2q \log q - \log q - 2 \log 2 + o(1)$ [with the
COL-UNCOL transition at $l_{\rm s}(q) = 2q \log q -\log q -1 + o(1)$].
Technically the size of dominating clusters is found by maximizing
$\Sigma(s)+s$ over the $s$ interval on which $\Sigma(s)\ge 0$. For
$\alpha\in [\alpha_{\rm c}(k),\alpha_{\rm s}(k)]$, the maximum is
reached at $s_{\rm max}$, with $\Sigma(s_{\rm max})= 0$ yielding $\phi
= s_{\rm max}$. It turns out that the solutions are comprised within a
finite number of clusters, with entropy $e^{Ns_{\rm max}+\Delta}$,
where $\Delta=\Theta(1)$.  The shifts $\Delta$ are asymptotically
distributed according to a Poisson point process of rate
$e^{-m(\alpha)\Delta}$ with $m(\alpha) = -\Sigma'(s_{\rm max})$.  This
leads to the Poisson Dirichlet distribution of weights discussed
above.  Finally, the entropy per variable $\phi$ is non-analytic at
$\alpha_{\rm c}(k)$.

Let us conclude by stressing two points. First, we avoided the $3$-SAT
and $3$-coloring cases.  These cases (as well as the $3$-coloring on
Erd\"os-R\'enyi graphs \cite{PalassiniMezardRivoire}) are particular
in that the dynamic transition point $\alpha_{\rm d}$ is determined by
a local instability (a Kesten-Stigum \cite{Kesten-Stigum} condition,
see also \cite{BiroliMonassonWeigt}), yielding $\alpha_{\rm d}(3)
\approx 3.86$ and $l_{\rm d}(3) = 6$ (the case $l=5$, $q=3$ being marginal).  
Related to this is the fact
that $\alpha_{\rm c}=\alpha_{\rm d}$: throughout the clustered phase,
the measure is dominated by a few large clusters (technically,
$\Sigma(s_*)<0$ for all $\alpha>\alpha_{\rm d}$).  Second, we did not
check the `local stability' of the 1RSB calculation. By analogy with
\cite{MontanariRicci}, we expect that an instability can modify the
curve $\Sigma(s)$ but not the values of $\alpha_{\rm d}$ and
$\alpha_{\rm c}$.

%
%
\subsection{Algorithmic implications}
\label{sec:Discussion}

Two message passing algorithms were studied extensively on random
$k$-SAT: belief propagation (BP) and survey propagation (SP) (mixed
strategies were also considered in \cite{MPEmpirical,Maneva}).  A BP
message $\nu_{u\to v}(x)$ between nodes $u$ and $v$ on the factor
graph is usually interpreted as the marginal distribution of $x_u$ (or
$x_v$) in a modified graphical model. An SP message is instead a
distribution over such marginals $P_{u\to v}(\nu)$.  The empirical
superiority of SP is usually attributed to the existence of clusters
\cite{MarcGiorgioRiccardo}: the distribution $P_{u\to v}(\nu)$ is a
`survey' of the marginal distribution of $x_u$ over the clusters.  As
a consequence, according to the standard wisdom, SP should outperform
BP for $\alpha>\alpha_{\rm d}(k)$.

This picture has however several problems. Let us list two of
them. First, it seems that essentially local algorithms (such as
message passing ones) should be sensitive only to correlations among
finite subsets of the variables\footnote{This paradox was noticed
independently by Dimitris Achlioptas (personal communication).}, and
these remain bounded up to the condensation transition.  Recall in
fact that the extremality condition in \eqref{eq:Correlation} involves
a number of variables unbounded in $N$, while the weaker in
\eqref{eq:Correlation2} is satisfied up to $\alpha_{\rm c}(k)$.
 
Secondly, it would be meaningful to weight uniformly the solutions
when computing the surveys $P_{u\to v}(\nu)$.  In the cavity method
jargon, this corresponds to using a 1RSB Parisi parameter $r=1$
instead of $r=0$ as is done in \cite{MarcGiorgioRiccardo}.  It is a
simple algebraic fact of the cavity formalism that for $r=1$ the means
of the SP surveys satisfy the BP equations. Since the means are the
most important statistics used by SP to find a solution, BP should
perform roughly as SP.
\begin{figure}[t]
\center{\includegraphics[width=0.85\linewidth]{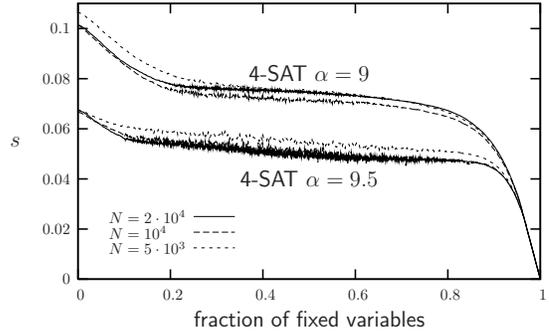}}
\caption{Performance of BP heuristics on random $4$-SAT formulae.  The
  residual entropy per spin $N^{-1}\log Z$ (here we estimate it within
  Bethe approximation) as a function of the fraction of fixed
  variables.  $t_\text{max}=20$ in these experiments.
\label{fig:Experiment}}
\end{figure}
Both arguments suggest that BP should perform well up to the
condensation point $\alpha_{\rm c}(k)$. We tested this conclusion on
$4$-SAT at $\alpha=9.5 \in(\alpha_{\rm d}(4),\alpha_{\rm c}(4))$,
through the following numerical experiment,
cf.~Fig.~\ref{fig:Experiment}.  $(i)$ Run BP for $t_{\rm max}$
iterations.  $(ii)$ Compute the BP estimates $\nu_i(x)$ for the
single bit marginals and choose the one with largest bias.  $(iii)$
Fix $x_i= 0$ or $1$ with probabilities $\nu_i(0)$, $\nu_i(1)$.  $(iv)$
Reduce the formula accordingly (i.e.\ eliminate the constraints
satisfied by the assignment of $x_i$ and reduce the ones violated).
This cycle is repeated until a solution is found or a contradiction is
encountered.  If the marginals $\nu_i(\,\cdot\,)$ were correct, this
procedure would provide a satisfying assignment sampled uniformly from
$\mu(\,\cdot\,)$.  In fact we found a solution with finite probability
(roughly $0.4$), despite the fact that $\alpha>\alpha_{\rm d}(4)$. The
experiment was repeated at $\alpha=9$ with a similar fraction of
successes (more data on the success probability will be reported in
\cite{nostro}).

Above the condensation transition, correlations become too strong and
the BP fixed point no longer describes the measure $\mu$.  Indeed the
same algorithm proved unsuccessful at $\alpha=9.7\in(\alpha_{\rm
c}(4),\alpha_{\rm s}(4))$.  As mentioned above, SP can be regarded as
an inference algorithm in a modified graphical model that weights
preferentially small clusters.  More precisely, it selects clusters of
size $e^{N\bar{s}}$ with $\bar{s}$ maximizing the complexity
$\Sigma(s)$.  With respect to the new measure, the weak correlation
condition in \eqref{eq:Correlation2} still holds and allows to perform
inference by message passing.

Within the cavity formalism, the optimal choice would be to take
$r\approx m(\alpha)\in [0,1)$.  Any parameter corresponding to a
non-negative complexity $r\in[0,m(\alpha)]$ should however give good
results.  SP corresponds to the choice $r=0$ that has some definite
computational advantages, since messages have a compact representation
in this case (they are real numbers).
%
%
\section{Cavity formalism, tree reconstruction and SP}
\label{sec:CavityFormalism}

This Section provides some technical elements of our computation.  The
reader not familiar with this topic is invited to further consult
Refs.~\cite{MarcGiorgioRiccardo,Mertens,PalassiniMezardRivoire,Revisited}
for a more extensive introduction. The expert reader will find a new
derivation, and some hints of how we overcame technical difficulties.
A detailed account shall be given in \cite{nostro,ours}.

On a tree factor graph, the marginals of $\mu(\,\cdot\,)$,
Eq.~\eqref{eq:Measure} can be computed recursively. The edge of the
factor graph from variable node $i$ to constraint node $a$
(respectively from $a$ to $i$) carries ``message'' $\oeta_{i \to a}$
($\onu_{a \to i}$), a probability measure on $\cX$ defined as the
marginal of $x_i$ in the modified graphical model obtained by deleting
constraint node $a$ ({\it resp}. all constraint nodes around $i$ apart
from $a$). The messages are determined by the equations
\begin{eqnarray}
\!\!
\oeta_{i\to a}(x_i)\!\!\! &=&\!\!\! \frac{1}{z_{i\to a}(\{\onu_{b\to i}\})} 
\prod_{b \in \di \setminus a}\onu_{b\to i}(x_i)\, ,\label{eq:VtoCeq}\\
\!\!\onu_{a\to i}(x_i)\!\!\! &=&\!\!\! 
\frac{1}{\widehat{z}_{a\to i}(\{\oeta_{j\to a}\})}
\sum_{\ux_{\da\setminus i}}\!\!\psi_a(\ux_{\da})\!\!\!
\prod_{j\in \da\setminus i}\!\!\oeta_{j\to a}(x_j)\, ,\label{eq:CtoVeq}
\end{eqnarray}
where $\partial u$ is the set of nodes adjacent to $u$, $\setminus$
denotes the set subtraction operation, and $\ux_A = \{x_j:j\in A\}$.
These are just the BP equations for the model \eqref{eq:Measure}.  The
constants $z_{i\to a}$, $\widehat{z}_{a\to i}$ are uniquely determined
from the normalization conditions $\sum_{x_i}\oeta_{i\to a}(x_i)=
\sum_{x_i}\onu_{a\to i}(x_i)=1$.  In the following we refer to these
equations by introducing functions $f_{i\to a}(\,\cdot\,)$, $f_{a\to
i}(\,\cdot\,)$ such that
\begin{equation}
\oeta_{i\to a} = 
f_{i \to a}(\{\onu_{b \to i} \}_{b \in \partial i \setminus a } ) \ , \ \
\onu_{a\to i} = 
f_{a \to i}(\{\oeta_{j \to a} \}_{j \in \partial a \setminus i } ) \ ,
\label{eq_recurs1}
\end{equation}
 The marginals of $\mu$ are
then computed from the solution of these equations. For instance
$\mu(x_i)$ is a function of the messages $\onu_{a \to i}$ from
neighboring function nodes.

The log-number of solutions, $\log Z$, can be expressed as a sum
of contributions which are local functions of the messages that 
solve Eqs.~\eqref{eq:VtoCeq}, \eqref{eq:CtoVeq} 
\begin{align}
\log Z =& \sum_a \log z_a(\{\oeta_{i \to a}\}) +
\sum_i \log z_i(\{\onu_{a \to i}\}) + \nonumber \\
& - \sum_{(ai)} \log z_{ai}(\oeta_{i \to a},\onu_{a \to i}) 
\label{eq_entropy}
\end{align}
where the last sum is over undirected edges in the factor graph and
\begin{align*}
z_a & \equiv \sum_{\ux_{\da}}\psi_a(\ux_{\da})\prod_{i\in\da}
\oeta_{i\to a}(x_i)\, ,\\
z_i & \equiv \sum_{x_i}\prod_{a\in\di}\onu_{a\to i}(x_i) \, , \qquad
z_{ai} \equiv \sum_{x_i}\oeta_{i\to a}(x_i)\onu_{a\to i}(x_i)\, .
\end{align*}
Each term $z$ gives the change in the number of solutions when merging
different subtrees (for instance $\log z_{i}$ is the change in entropy
when the subtrees around $i$ are glued together). This expression
coincides with the Bethe free-energy \cite{Yedidia} as expressed in
terms of messages.

In order to move from trees to loopy graphs, we first consider an
intermediate step in which the factor graph is still a tree but a
subset of the variables, $\uxb = \{x_j:j\in \bo\}$ is fixed.  We are
therefore replacing the measure $\mu(\,\cdot\,)$,
cf. Eq.~\eqref{eq:Measure}, with the conditional one
$\mu(\,\cdot\,|\uxb)$. In physics terms, the variables in $\uxb$
specify a boundary condition.

Notice that the measure $\mu(\,\cdot\,|\uxb)$ still factorizes
according to (a subgraph of) the original factor graph. As a
consequence, the conditional marginals $\mu(x_i|\uxb)$ can be computed
along the same lines as above.  The messages $\eta_{i \to a}^{\uxb}$
and $\nu_{a \to i}^{\uxb}$ obey Eqs.~\eqref{eq_recurs1}, with an
appropriate boundary condition for messages from $\bo$. Also, the
number of solutions that take values $\uxb$ on $j\in B$ (call it
$Z(\uxb)$) can be computed using Eq.~\eqref{eq_entropy}.

Next, we want to consider the boundary variables themselves as random
variables. More precisely, given $r\in{\mathbb R}$, we let the
boundary to be $\uxb$ with probability
\begin{equation}
\widetilde{\mu}(\uxb) = \frac{Z(\uxb)^r}{\Z (r) } \ ,\label{eq:MuTilde}
\end{equation}
where $\Z(r)$ enforces the normalization
$\sum_{\uxb}\widetilde{\mu}(\uxb)=1$. Define $P_{i \to a}(\eta)$ as
the probability density of $\eta_{i \to a}^{\uxb}$ when $\uxb$ is
drawn from $\widetilde{\mu}$, and similarly $Q_{a \to i}(\nu)$. One
can show that Eq.~\eqref{eq:VtoCeq} implies the following relation
between messages distributions
\begin{equation}
P_{i \to a}(\eta)=
\frac{1}{\Z_{i \to a}} \int\!\!\!\prod_{b \in \partial i \setminus a }
\!\!\de Q_{b \to i}(\nu_{b})\,
\delta[\eta -  f_{i \to a}(\{\nu_{b} \})]\,
z_{i \to a}(\{\nu_{b} \})^r,
\label{eq_recurs2}
\end{equation}
where $f_{i \to a}$ is the function defined in Eq.~\eqref{eq_recurs1},
$z_{i \to a}$ is determined by Eq.~\eqref{eq:VtoCeq}, and $\Z_{i \to
a}$ is a normalization. A similar equation holds for $Q_{a \to
i}(\nu)$.  These coincide with the ``1RSB equations'' with Parisi
parameter $r$.  Survey propagation (SP) corresponds to a particular
parameterization of Eq.~\eqref{eq_recurs2} (and the analogous one
expressing $Q_{a\to i}$ in terms of the $P$'s) valid for $r=0$.

The log-partition function $\Phi(r) = \log \Z(r)$ admits an expression
that is analogous to Eq.~\eqref{eq_entropy},
\begin{align}
\log \Z(r) = & \sum_a \log \Z_a(\{P_{i \to a}\}) +
\sum_i \log \Z_i(\{Q_{a \to i}\}) + \nonumber \\
& -\sum_{ai} \log \Z_{ai}(P_{i \to a},Q_{a \to i})
\label{eq_potential}
\end{align}
where the `shifts' $\Z(\cdots)$ are defined through moments of order
$r$ of the $z$'s, and sums run over vertices not in $\bo$.  For
instance $\Z_{ai}$ is the expectation of $z_{ai}(\eta,\nu)^r$ when
$\eta$, $\nu$ are independent random variables with distribution
(respectively) $P_{i \to a}$ and $Q_{a \to i}$.  The (Shannon) entropy
of the distribution $\widetilde{\mu}$ is given by $\Sigma(r)=\Phi(r)-r
\Phi'(r)$.

As mentioned, the above derivation holds for tree factor graphs.
Nevertheless, the local recursion equations \eqref{eq_recurs1},
\eqref{eq_recurs2} can be used as an heuristics on loopy factor graphs
as well.  Further, although we justified Eq.~\eqref{eq_recurs2}
through the introduction of a random boundary condition $\uxb$, we can
take $\bo=\emptyset$ and still look for non-degenerate solutions of
such equations.

Starting from an arbitrary initialization of the messages, the
recursions are iterated until an approximate fixed point is reached.
After convergence, the distributions $P_{i\to a}$, $Q_{a\to i}$ can be
used to evaluate the potential $\Phi(r)$,
cf.~Eq.~\eqref{eq_potential}.  From this we compute the complexity
function $\Sigma(r)\equiv \Phi(r)-r \Phi'(r)$, that gives access to
the decomposition of $\mu(\,\cdot\,)$ in pure states.  More precisely,
$\Sigma(r)$ is the exponential growth rate of the number of states
with internal entropy $s=\Phi'(r)$.  This is how curves such as in
Fig.~\ref{fig:Complexity} are traced.

In practice it can be convenient to consider the distributions of
messages $P_{i\to a}$, $Q_{a\to i}$ with respect to the graph
realization.  This approach is sometimes referred to as `density
evolution' in coding theory.  If one consider a uniformly random
directed edge $i\to a$ (or $a\to i$) in a rCSP instance, the
corresponding message will be a random variable.  After $t$ parallel
updates according to Eq.~\eqref{eq_recurs2}, the message distribution
converges (in the $N\to\infty$ limit) to a well defined law ${\cal
P}_t$ (for variable to constraint messages) or ${\cal Q}_t$ (for
constraint to variable).  As $t\to\infty$, these converge to a fixed
point ${\cal P}$, ${\cal Q}$ that satisfy the distributional
equivalent of Eq.~\eqref{eq_recurs2}.

To be definite, let us consider the case of graph coloring.  Since the
compatibility functions are pairwise in this case (i.e.  $k=2$ in
Eq.~\eqref{eq:Measure}), the constraint-to-variable messages can be
eliminated and Eq.~\eqref{eq_recurs2} takes the form
\begin{align*}
P_{i \to j}(\eta)\propto
\int\!\!\!\prod_{l \in \partial i \setminus j}
\!\!\de P_{l \to i}(\eta_{l})\,
\delta\left[\eta - f(\{\eta_l\})\right] 
z(\{\eta_l \})^r\, ,
\end{align*}
where $f$ is defined by $\eta(x) = z^{-1}\prod_{l} 1-\eta_l(x)$ and
$z$ by normalization. The distribution of $P_{i\to j}$ is then assumed
to satisfy a distributional version of the last equation.  In the
special case of random regular graphs, a solution is obtained by
assuming that $P_{i\to j}$ is indeed independent of the graph
realization and of $i,j$. One has therefore \emph{simply} to set
$P_{i\to j} = P$ in the above and solve it for $P$.

In general, finding messages distributions ${\cal P}$, ${\cal Q}$ that
satisfy the distributional version of Eq.~\eqref{eq_recurs2} is an
extremely challenging task, even numerically.  We adopted the
population dynamics method \cite{Revisited} which consists in
representing the distributions by samples (this is closely related to
particle filters in statistics).  For instance, one represents ${\cal
P}$ by a sample of $P$'s, each encoded as a list of $\eta$'s.  Since
computer memory drastically limits the samples size, and thus the
precision of the results, we worked in two directions: $(1)$ We
analytically solved the distributional equations for large $k$ (in the
case of $k$-SAT) or $q$ ($q$-coloring); $(2)$ We identified and
exploited simplifications arising for special values of $r$.

Let us briefly discuss point $(2)$. Simplifications emerge for $r=0$
and $r=1$. The first case correspond to SP:
Refs.~\cite{MarcGiorgioRiccardo,Mertens} showed how to compute
efficiently $\Sigma(r=0)$ through population dynamics.  Building on
this, we could show that the clusters internal entropy $s(r=0)$ can be
computed at a small supplementary cost (see \cite{nostro}).

The value $r=1$ corresponds instead to the `tree reconstruction'
problem \cite{TreeRec}: In this case $\widetilde{\mu}(\uxb)$,
cf. Eq.~\eqref{eq:MuTilde}, coincides with the marginal of $\mu$.
Averaging Eq.~\eqref{eq_recurs2} (and the analogous one for $Q_{a\to
i}$) one obtains the BP equations \eqref{eq:VtoCeq},
\eqref{eq:CtoVeq}, e.g. $\int \de P_{i \to a}(\eta)\ \eta = \oeta_{i
\to a}$.  These remark can be used to show that the constrained
averages
\begin{equation}
\overline{P}(\eta , \oeta ) = \int \de
{\cal P}[P] \ P(\eta) \ 
\delta \left( \oeta - \int \de P(\eta') \eta' \right) \ ,
\label{eq:M1Recursion}
\end{equation}
and $\overline{Q}(\nu,\onu)$ (defined analogously)
satisfy closed equations which are much easier to solve
numerically.
%
%
\vspace{-0.3cm}

\begin{acknowledgments}
We are grateful to J. Kurchan and M. M\'ezard for stimulating
discussions.  This work has been partially supported by the European
Commission under contracts EVERGROW and STIPCO.
\end{acknowledgments}

%
%

%
\end{article}


\begin{thebibliography}{99} 
%
\bibitem{GaJo} Garey MR and Johnson DS (1979) {\it Computers and
    Intractability: A Guide to the Theory of NP-Completenes} (Freeman,
    New York).
%
\bibitem{Factor} Kschischang FR, Frey BJ and Loeliger H-A (2001) {\it
    IEEE~Trans.~Inform.~Theory} {\bf 47}, 498-519.
%
\bibitem{FrancoPaull} Franco J and Paull M (1983) {\it Discrete
  Appl.~Math} {\bf 5}, 77-87.
%
\bibitem{WalkSAT} Selman B, Kautz HA and Cohen B (1994) {\it Proc.~of
    AAAI-94, Seattle}.
%
\bibitem{Myopic} Achlioptas D and Sorkin GB (2000) {\it Proc.~of the
    Annual Symposium on the Foundations of Computer Science, Redondo
    Beach, CA}.
%
\bibitem{MarcGiorgioRiccardo} M\'ezard M, Parisi G and Zecchina R
  (2002) {\it Science} {\bf 297}, 812-815.
%
\bibitem{BayatiEtAl} Bayati M, Shah D and Sharma M (2006) {\it
    Proc.~of International Symposium on Information Theory, Seattle}.
%
\bibitem{GamarnikColoring} Gamarnik D and Bandyopadhyay A (2006) {\it
    Proc.~of the Symposium on Discrete Algorithms, Miami}.
%
\bibitem{MontanariShah} Montanari A and Shah D (2007) {\it Proc.~of
    the Symposium on Discrete Algorithms, New Orleans}.
%
\bibitem{Friedgut} Friedgut E (1999) {\it J.~Amer.~Math.~Soc.} {\bf
    12}, 1017-1054.
%
\bibitem{Mertens} Mertens S, M\'ezard M and Zecchina R (2006) {\it
    Rand.~Struct.~and Alg.} {\bf 28}, 340-373.
%
\bibitem{AchlioptasNaorPeres} Achlioptas D, Naor A and Peres Y
    (2005) {\it Nature} {\bf 435}, 759-764.
%
\bibitem{ColTrieste} Mulet R, Pagnani A, Weigt M and Zecchina R (2002)
  {\it Phys.~Rev.~Lett.} {\bf 89}, 268701.
%
\bibitem{KrzakalaPagnaniWeigt} Krzakala F, Pagnani A and Weigt M
  (2004) {\it Phys.~Rev.~E} {\bf 70}, 046705.
%
\bibitem{Luczak} Luczak T (1991) {\it Combinatorica} {\bf 11}, 45-54.
%
\bibitem{MontanariMezard} M\'ezard M and Montanari A (2006)
  {\it J.~Stat.~Phys.} {\bf 124}, 1317-1350.
%
\bibitem{Georgii} Georgii HO (1988) {\it Gibbs Measures and Phase
  Transitions} (De Gruyter, Berlin).
%
\bibitem{Jonasson} Jonasson J (2002) {\it Stat.~and Prob.~Lett.} {\bf
  57}, 243-248.
%
\bibitem{MPEmpirical} Aurell E, Gordon U and Kirkpatrick S (2004) {\it
    Proc.~of Neural Information Processing Symposium, Vancouver}.
%
\bibitem{Maneva} Maneva EN, Mossel E and Wainwright MJ (2005) {\it
  Proc.~of the Symposium on Discrete Algorithms, Vancouver}.
%
\bibitem{BiroliMonassonWeigt} Biroli G, Monasson R and Weigt M (2000)
    {\it Eur.~Phys.~J.~B} {\bf 14}, 551-568.
%
\bibitem{MezardMoraZecchina} M\'ezard M, Mora T and Zecchina R (2005)
    {\it Phys.~Rev.~Lett.} {\bf 94}, 197205.
%
\bibitem{AchlioptasRicci} Achlioptas D and Ricci-Tersenghi F (2006)
    {\it Proc.~of the Symposyum on the Theory of Computing,
    Washington}.
%
\bibitem{AlfredoRiccardo} Braunstein A and Zecchina R (2004) {\it
    J.~Stat.~Mech.}, P06007.
%
\bibitem{PalassiniMezardRivoire} M\'ezard M, Palassini M and Rivoire O
  (2005) {\it Phys.~Rev.~Lett.} {\bf 95}, 200202.
%
\bibitem{Ruelle} Ruelle D (1987) {\it Commun.~Math.~Phys.} {\bf 108},
  225-239.
%
\bibitem{TalagrandPspin} Talagrand M (2000) {\it
  C.~R.~Acad.~Sci.~Paris}, Ser.~I {\bf 331}, 939-942.
%
\bibitem{OurGibbs} Montanari A and Semerjian G, {\it
In preparation.}
%
\bibitem{Kesten-Stigum} Kesten H and Stigum BP (1966) {\it
  Ann.~Math.~Statist.} {\bf 37}, 1463-1481.  Kesten H and Stigum BP
  (1966) {\it J.~Math.~Anal.~Appl.} {\bf 17}, 309-338.
%
\bibitem{MontanariRicci} Montanari A and Ricci-Tersenghi F (2004)
  {\it Phys.~Rev.~B} {\bf 70}, 134406.
%
\bibitem{nostro} Montanari A, Ricci-Tersenghi F and Semerjian G, {\it
    In preparation.}
%
\bibitem{Revisited} M\'ezard M and Parisi G (2001) {\it
  Eur.~Phys.~J.~B} {\bf 20}, 217-233.
%
\bibitem{ours} Krzakala F and Zdeborov{\'a} L, {\it In preparation.}
%
\bibitem{Yedidia} Yedidia J, Freeman WT and Weiss Y (2005) 
{\it  IEEE~Trans.~Inform.~Theory}  {\bf 51}, 2282-2312.
%
\bibitem{TreeRec} Mossel E and Peres Y (2003) {\it Ann. Appl. Probab.}
  {\bf 13}, 817-844.
%
\end{thebibliography}
\end{document}